\documentclass{pasa}%

\usepackage{graphicx}
\usepackage{booktabs,caption}
\usepackage[flushleft]{threeparttable}

\title[Bright $z\sim5$ quasars from SkyMapper]{Discovery of two bright $z\sim5$ quasars with SkyMapper, Pan-STARRS1 and WISE}

\author[Li et al.]{Zefeng Li$^{1,2}$, Christian Wolf$^{1,2}$, Fuyan Bian$^{1,3}$, Christopher A. Onken$^{1,2}$, Brian P. Schmidt$^{1,2}$, Patrick Tisserand$^{1,4}$, Noura Alonzi$^{5,6}$ and Wei Jeat Hon$^{5}$ 
\affil{$^1$Research School of Astronomy and Astrophysics, Australian National University, Canberra, ACT 2611, Australia}%
\affil{$^2$ARC Centre of Excellence for All-sky Astrophysics (CAASTRO)}
\affil{$^3$European Southern Observatory (ESO), Vitacura, Chile}
\affil{$^4$Sorbonne Universit\'{e}s, UPMC Univ Paris 6 et CNRS, Institut d'Astrophysique de Paris, 98 bis bd Arago, F-75014 Paris, France}
\affil{$^5$School of Physics, University of Melbourne, Parkville, Victoria 3010, Australia}
\affil{$^6$Department of Physics and Astronomy, King Saud University, Riyadh 11451, Saudi Arabia}
}%

\jid{PASA}
\doi{10.1017/pas.\the\year.xxx}
\jyear{\the\year}

\usepackage{aas_macros}
\usepackage{hyperref} 
\hypersetup{colorlinks,citecolor=blue,linkcolor=blue,urlcolor=blue}

\begin{document}

\begin{frontmatter}
\maketitle

\begin{abstract}
We present a search for bright $z\sim5$ quasars using imaging data from SkyMapper Southern Survey, Pan-STARRS1 and the Wide-field Infrared Survey Explorer (WISE). We select two sets of candidates using WISE with optical bands from SkyMapper and alternatively from Pan-STARRS1, limited to a magnitude of $i<18.2$. We follow up several candidates with spectroscopy and find that the four candidates common to both lists are quasars, while others turned out to be cool stars. Two of the four quasars, SMSS J013539.27-212628.4 at $z=4.86$ and SMSS J093032.58-221207.7 at $z=4.94$, are new discoveries and ranked among the dozen brightest known $z>4.5$ QSOs in the $i$-band.
\end{abstract}

\begin{keywords}
galaxies: active - galaxies:high-redshift - quasars: general
\end{keywords}
\end{frontmatter}

\section{INTRODUCTION }\label{sec:intro}

Quasars were first discovered by \citet{Schmidt63} and are the most optically luminous variety of active galactic nuclei (AGN). Their central engines are powered by massive accretion of matter onto supermassive black holes (SMBHs). At high redshift, quasars provide a powerful tool to study the formation and growth history of SMBHs \citep[e.g.][]{Mortlock11,Wu15}, probe the progress of cosmic reionisation \citep[e.g.][]{Fan06a, Fan06b}, and constrain the metal enrichment and dust production in the early epoch of the universe \citep[e.g.][]{Jiang16}. Bright quasars at high redshift are particularly attractive finds; they are powered by the most massive black holes; they are potentially one of the major contributors to reionisation; and they are the brightest beacons, highlighting the chemical evolution of the universe most effectively \citep{Ryan-Weber09}, so that it can be observed with good signal-to-noise.

Bright quasars at redshifts of $z>4.5$ are exceedingly rare objects as less than a dozen of them are known with a magnitude of $i<18$. They are generally faint as the brightest known object has $i=17.25$, and they are optically prominent only in the $i$- and $z$-band as intergalactic Hydrogen absorption removes most of their short-wavelength emission. Finding them requires large surveys and so it is no surprise that the first $z>5$ quasars were found in the large-area Sloan Digital Sky Survey \citep[SDSS,][]{Fan99,Fan01,Schneider10}. Quasars at $z>4.5$ are routinely discovered in varieties of wide-field surveys, including the Canada-France High-z Quasar Survey \citep[CFHQS,][]{Willott07}, the UKIRT Infrared Deep Sky Survey \citep[UKIDSS,][]{Venemans07, Yang17}, the Panoramic Survey Telescope and Rapid Response System \citep[Pan-STARRS,][]{Banados16}, the Dark Energy Survey \citep[DES,][]{Reed15}, and the Dark Energy Camera Legacy Survey \citep[DECaLS,][]{Wang17}. Nevertheless, bright $z>4.5$ quasars are difficult to find because of their intrinsic rareness and severe contamination by cool stars with broadly similar colours. 

The $r-i/i-z$ colour-colour diagram is a widely used method to select $z\sim5$ quasars \citep{Zheng00, Fan01, Chiu05}. Recently, large-area infrared surveys made it possible to select quasar candidates with less contamination from cool stars. \cite{McGreer13} observed 92 quasar candidates selected by combining the SDSS imaging with addition $J$-band imaging. Seventy-three out of 92 candidates, i.e. 80\%, were indeed confirmed to be $4.7 \le z \le 5.1$ quasars. Unfortunately, this method is limited to a narrow redshift range \citep{McGreer13}. \cite{Wang16} identified 72 new $z\sim5$ quasars by combining SDSS data with photometry from the Wide-field Infrared Survey Explorer \citep[WISE,][]{Wright10}.

The SDSS only covers a large portion of the Northern sky, although the Pan-STARRS1 survey \citep[PS1,][]{Kaiser02, Kaiser10, Chambers11} has extended coverage of 1.5 hemispheres to a declination of $-30\deg$ South, but a digital atlas of the full Southern hemisphere has been missing. The SkyMapper Southern Survey aims at plugging that gap, and is a full hemispheric imaging survey carried out by the SkyMapper telescope at Siding Spring Observatory in New South Wales, Australia. The first data release \citep[DR1,][]{Wolf18}\footnote{http://skymapper.anu.edu.au} covers an area of over 20,200~$\mathrm{deg^2}$ of Southern sky and $\sim$285 million objects in the six SkyMapper passbands $uvgriz$. This first release includes only data from the SkyMapper Shallow Survey and has $10\sigma$-limiting AB magnitudes of $\sim 18$ in all passbands.

This paper presents our first step in searching for bright high-redshift QSOs in the southern sky and reports the discovery of two new bright $i\sim 18$ quasars at $z\sim5$. Throughout the paper, WISE magnitudes are used in the Vega system as is customary, while all optical magnitudes are in the AB system \citep{Oke83}. 
%We adopt a standard $\mathrm{\Lambda CDM}$ cosmology with parameters $\mathrm{H_0} = 70 \mathrm{km s^{-1} Mpc^{-1}}$, and density parameters $\mathrm{\Omega_{M}} =0.3$ and $\mathrm{\Omega_{\Lambda}} =0.7$.

\section{QUASAR SELECTION}\label{Section_selection}

At $z\sim5$, most quasars are very faint in $g$-band, because Lyman limit systems (LLSs), which are optically thick to the continuum radiation from the quasar \citep{Fan99}, are redshifted into the band. Also, Lyman series absorption systems begin to dominate in the $r$-band and Ly$\alpha$ emission moves to the $i$-band, leading to a red $r-i$ colour \citep{Wang16}. Thus, in previous studies the $r-i/i-z$ diagram has often been used to select $z\sim5$ quasar candidates \citep{Fan99, Richards02, McGreer13}. 

SkyMapper's first data release includes only Shallow Survey data and is thus only complete to $\sim 18$~mag in the $riz$-bands. Prior to our work, only ten quasars were known with $i_{\rm SDSS}<18.2$. While we searched for QSOs with $i_{\rm SM}<18.2$, we realised that the $r$-band in SkyMapper is not quite deep enough to search for very red $r-i$ colours with sufficiently high significance, and as a result we get too many candidates that are expected to be faint cool stars. Hence, we draw on Pan-STARRS1 for deeper $g$- and $r$-band photometry.

\begin{figure}
\begin{center}
\includegraphics[width=\columnwidth]{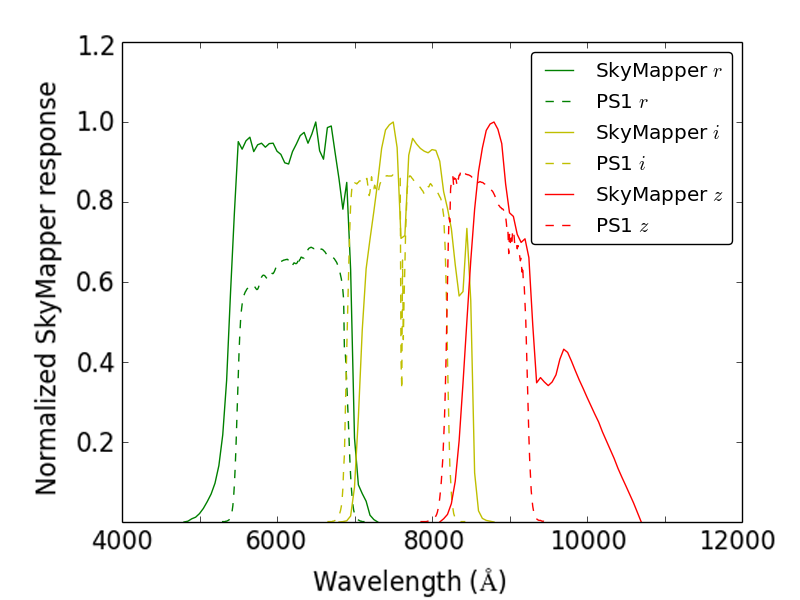}
\caption{Filter response in SkyMapper (normalised) and Pan-STARRS1 $riz$-bands. While the $r$-band filters are very similar, the SkyMapper version of $i$ and $z$ reach further into the red; however, SkyMapper has no equivalent to the Pan-STARRS1 $y$-band.}
\label{filters}
\end{center}
\end{figure}

\begin{figure*}
\begin{center}
\includegraphics[width=1.0\textwidth]{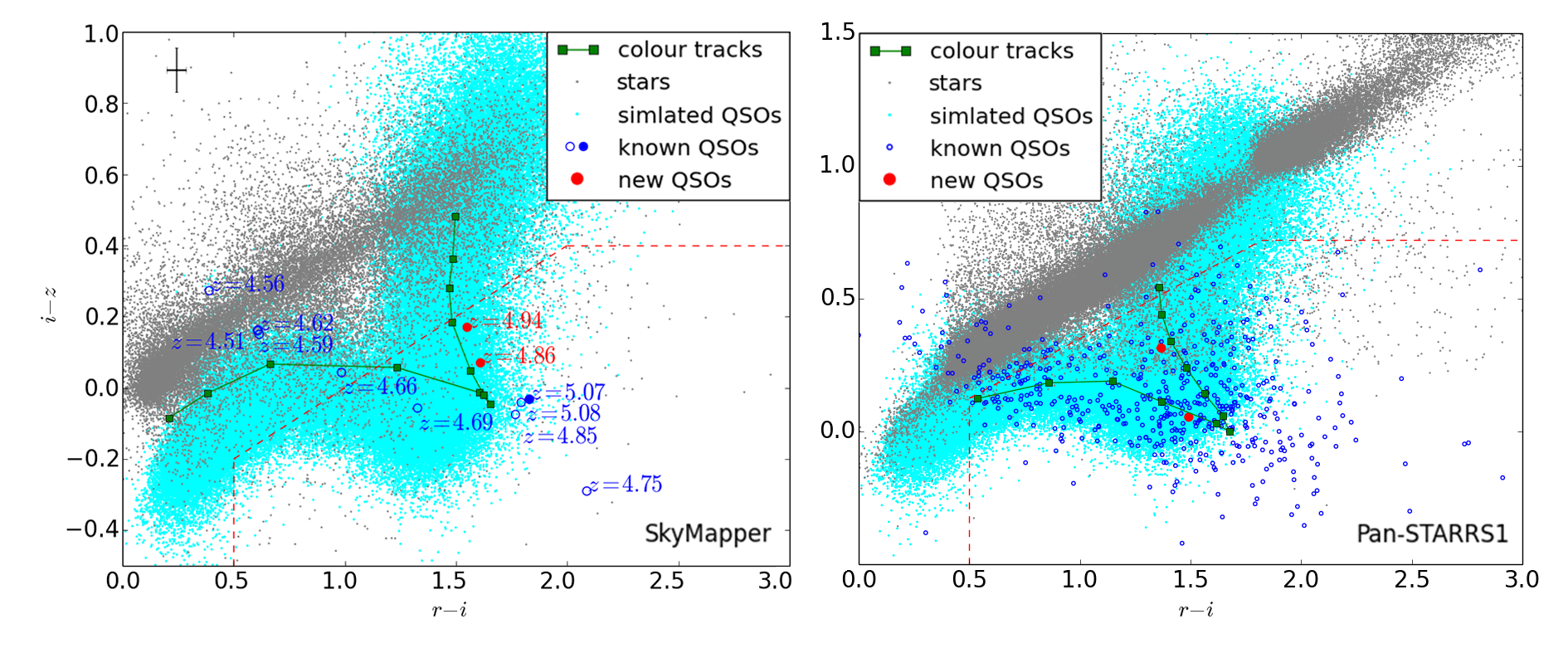}
\caption{Comparison of the $i-z$ vs. $r-i$ colour-colour diagram for SkyMapper (left panel, $r_{\rm PS1}, i_{\rm SM}$ and $z_{\rm SM}$) and Pan-STARRS1 (right panel, $r_{\rm PS1}, i_{\rm PS1}$ and $z_{\rm PS1}$). In both panels simulated quasars are derived from \cite{McGreer13} and the colour tracks correspond to the quasar template of \cite{VandenBerk01} for redshifts from $z$=4.4 to $z$=5.5 in steps of $\Delta z$=0.1. The red dashed lines represent the selection criteria. In the left panel, the blue circles are the only 10 known $z\sim5$ quasars detected in SkyMapper survey for now, and among them the blue solid point is the quasar discovered by \cite{Wang16}. Typical photometric errors are shown in the upper-left corner in the SkyMapper panel, while they are negligible in the Pan-STARRS1 version.}
\label{Figure_optical_criteria}
\end{center}
\end{figure*}

The $riz$-passbands in SkyMapper and Pan-STARRS1 are similar in wavelength coverage (see Fig.~\ref{filters}): the $r$-bands are nearly identical, but the $i$- and $z$-bands of SkyMapper reach longer wavelengths than the Pan-STARRS1 cousins; Pan-STARRS1 instead uses an extra $y$-band filter to cover the longest wavelengths in the CCD sensitivity range. The passband differences affect the $r-i/i-z$ colours of very cool stars, and in Fig.~\ref{Figure_optical_criteria} we show the $r-i/i-z$ diagram for both SkyMapper and Pan-STARRS1. Due to the shallow data of the SkyMapper DR1, the photometric errors on the $i$-band drive the $i-z$ colour a little higher than is the case in the PS1 DR1. This shortage will be overcome later in 2018, when the SkyMapper DR2 is released with expected limits beyond 20 mag in all filter bands. 

We note the different placement of the stellar locus in the colour-colour diagrams of the two surveys, and adjust selection cuts accordingly to avoid too much stellar contamination. We select candidates in the overlapping part of the sky surveyed by Pan-STARRS1 and SkyMapper twice, once using the SkyMapper $i$/$z$-filters and once using PS1 filters; we find that the two lists have only a modest number of candidates in common. One possible reason could be that Pan-STARRS1 does not observe its filters simultaneously, while SkyMapper does (within a 5-minute window in the Shallow Survey). We speculate that while SkyMapper DR1 has the noisier photometry, Pan-STARRS1 colours might be affected by variability when e.g. M star flares affect the photometry of just one band at the time of observation. However, a detailed analysis of such issues is beyond the scope of this paper.

Infrared photometry has long been considered useful for rejecting cool-star contaminants among high-z quasar candidates \citep{Pentericci03, McGreer13}. However, for a large-area search we would need nearly all-sky $J$-band data such as that provided by the 2 Micron All-Sky Survey \citep[2MASS,][]{2MASS}. Because of the relative shallowness of the 2MASS $J$-band data, we cannot directly duplicate the method reported by \cite{McGreer13}.

\begin{figure}
\begin{center}
\includegraphics[width=1.0\columnwidth]{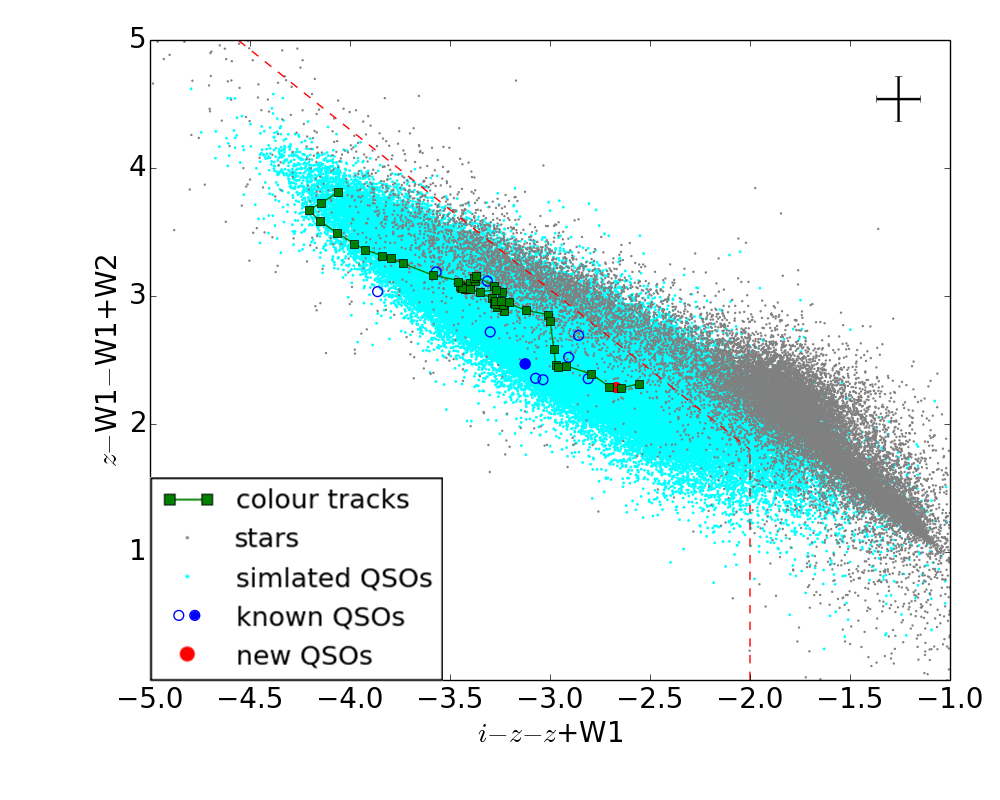}
\caption{The $i-z-z+$W1 vs. $z-$W1$-$W1$+$W2 colour-colour diagram. From left to right, the colour track is from $z=1.0$ to $z=5.5$ in steps of $\Delta z=0.1$. Typical photometric error bars are shown in the upper-right corner.}
\label{Figure_MIR_criteria}
\end{center}
\end{figure}

However, we do include mid-infrared (MIR) WISE\footnote{We used the WISE ALLWISE catalogue \citep{Cutri13}.} photometry in the selection, but explore an alternative approach to the use of the $z-$W1/W1$-$W2 diagram used by \citet{Wang16} in order to be more complete in the redshift range $z=[4.5,5]$. Thus, we adopt a new selection to cover a wider redshift range. We combine two colours to measure the curvature in the spectral energy distribution (SED). A star has typically a bump-shaped SED, and hence its $i-z$ colour is redder and its $z-$W1 colour is bluer than that of a quasar. As a result a star has a much redder $i-z-z+$W1 value than a quasar (see Fig.~\ref{Figure_MIR_criteria}). This MIR selection does not discriminate low from high redshifts, but is just intended to separate stars from quasars. 

We start by retrieving candidates for the two versions of $i/z$-filters using the TAP queries shown in the Appendix. In both versions of the query, the criteria are composed of three sections. The first section selects required columns and joins the SkyMapper master table with PS1 and WISE tables using the pre-calculated cross-match in the SkyMapper database. The second part requires that SkyMapper must not detect the object in $uvgr$-bands; reject clearly extended sources ($z_{\rm psf}-z_{\rm petro}$); require galactic latitude $|b|>20$; require an isolated source, where the nearest neighbour in SkyMapper is at least 10 arcsec away; various reliability flags indicate that measurements are clean; and that the source has reliable Pan-STARRS1 magnitudes. The third section contains the colour criteria, which differ between the two survey versions. Notably, a follow-up point-source confirmation for the Pan-STARRS1 version is done by checking iPSFMag-iKronMag<0.05 after being cross-matched with the Pan-STARRS1 StackObject database \citep{Farrow14}. We listed this part in square brackets, because it has to be separately retrieved via Pan-STARRS1 Casjob\footnote{http://mastweb.stsci.edu/ps1casjobs/} instead of in the TAP query.

The SkyMapper query returns 53 candidates. We eyeballed image cutouts of these 53 candidates and removed 3 candidates with suspicious features, while the remaining 50 objects form the SkyMapper $z\sim5$ quasar candidate list. The Pan-STARRS1 query returns 41 candidates and makes up the Pan-STARRS1 list. For our spectroscopic follow-up we concentrated on the overlap section of the two lists, which contained only four candidates simultaneously selected by the both surveys, two of which are already known $z>4.5$ QSOs. Afterwards, we followed up some of the remaining candidates from the two lists.

\section{SPECTROSCOPIC OBSERVATIONS}\label{Section_Spectroscopic_Observations}

For the spectroscopic follow-up observations we used the ANU 2.3m-telescope at Siding Spring Observatory in Australia with the Wide Field Spectrograph \citep[WiFeS,][]{Dopita07, Dopita10} for part of the nights on December 21 to 23, 2017, January 12 to 13, 2018 and April 11 to 13, 2018. For each candidate we obtained one 1800-second exposure with the R3000 grating, which gives a resolution of R=3000 at wavelengths between 5300~\AA \ and 9800~\AA. The WiFeS data were reduced using a Python-based pipeline, PyWiFeS \citep{Childress14}. The flux calibrations of all spectra were conducted from the standard stars observed on the same night.

\begin{figure}
\begin{center}
\includegraphics[width=1.0\columnwidth]{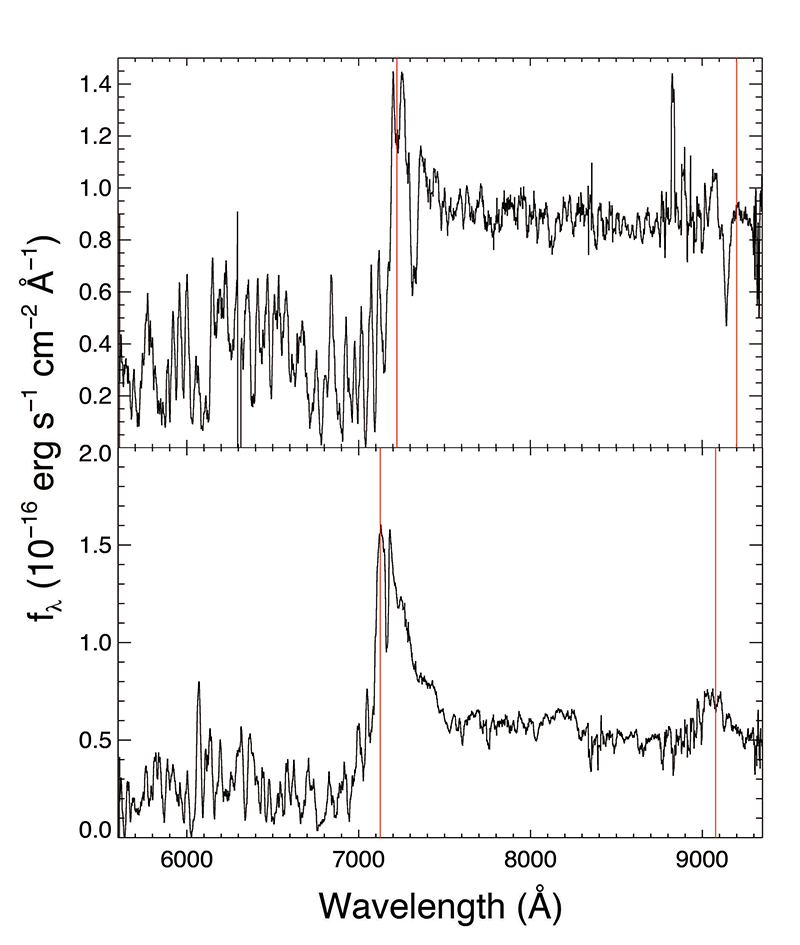}
\caption{Spectra of the two new bright $z\sim5$ quasars, J013539-212628 at $z=4.94$ (upper panel) and J093032-221208 at $z=4.86$ (lower panel). The red lines mark the expected positions of $\mathrm{Ly\alpha}$ (left) and $\mathrm{CIV}$ (right).}
\label{Figure_spectra}
\end{center}
\end{figure}

We observed 15 candidates from the two lists and classified 14 of the objects. Two of them are newly identified as bright quasars at $z=4.86$ and $z=4.94$, while one of them was accidentally observed by us and had been identified as a high-z quasar previously (J002526-014532). The remaining objects turn out to be cool stars. As a result all four candidates from the common candidate list are quasars at $z>4.5$, and two of these are newly identified in this work, while \cite{Wang16} already reported J002526-014532 and J211105-015604. The properties of all of our observed candidates are listed in Table 1, including the stars we verified. We do not claim that there are no more quasars to be found in the single-survey candidate lists, but we had no opportunity to follow these up. The spectra of the two newly identified quasars are shown in Fig.~\ref{Figure_spectra}.

\begin{table*}
\label{Table_properties}
\begin{center}
\caption{Properties of the spectroscopically verified candidates. The top three rows are quasars, followed by stars (nine rows of SkyMapper-only and two rows of Pan-STARRS1-only candidates). Errors of $i_{\rm PS1}$ and $z_{\rm PS1}$ are all typically 0.01.}
\begin{tabular}{cccccccr}
\hline Object & $z$ & $i_\mathrm{SM}$ & $z_\mathrm{SM}$ &
$r_\mathrm{PS1}$ & $i_\mathrm{PS1}$ & $z_\mathrm{PS1}$ & W1$-$W2 ~ \\
\hline
J013539.27-212628.4 & 4.94 & $18.00\pm0.08$ & $17.83\pm0.02$ &
$19.55\pm0.01$ & $18.18$ & $17.86$ & $0.68\pm0.04$\\ % $14.13\pm0.02$ & $13.45\pm0.03$ \\
J093032.58-221207.7 & 4.86 & $18.11\pm0.02$ & $18.04\pm0.02$ &
$19.72\pm0.01$ & $18.23$ & $18.17$ & $0.44\pm0.21$\\ % $15.45\pm0.10$ & $15.01\pm0.19$ \\
\hline
J002526.82-014532.7 & 5.07 & $17.89\pm0.06$ & $17.74\pm0.25$ &
$19.76\pm0.02$ & $18.13$ & $18.04$ & $0.62\pm0.08$\\ % $14.83\pm0.04$ & $14.21\pm0.07$ \\
\hline
J084734.65-044343.5 & star & $18.19\pm0.21$ & $18.00\pm0.05$ &
$19.90\pm0.02$ & $18.57$ & $17.96$ & $0.23\pm0.10$\\ % $14.89\pm0.03$ & $14.66\pm0.10$ \\
J091437.47-162248.1 & star  & $18.17\pm0.06$ & $18.05\pm0.05$ &
$19.79\pm0.01$ & $18.51$ & $17.92$ & $0.27\pm0.13$\\ % $15.50\pm0.06$ & $15.23\pm0.11$ \\
J093207.25-003728.3 & star  & $18.18\pm0.03$ & $18.15\pm0.03$ &
$19.92\pm0.03$ & $18.52$ & $17.89$ & $0.16\pm0.00$\\ % $15.39\pm0.07$ & $15.23\pm0.16$ \\
J095217.30-003712.5 & star  & $18.16\pm0.08$ & $18.00\pm0.05$ &
$19.79\pm0.02$ & $18.52$ & $17.94$ & $0.31\pm0.17$\\ % $15.30\pm0.04$ & $14.99\pm0.16$ \\
J110023.13-204211.3 & star  & $18.14\pm0.16$ & $17.88\pm0.09$ &
$19.91\pm0.01$ & $18.52$ & $17.90$ & $0.54\pm0.20$\\ % $15.35\pm0.10$ & $14.81\pm0.17$ \\
J114146.56-261634.7 & star  & $18.17\pm0.04$ & $18.26\pm0.08$ &
$19.78\pm0.15$ & $18.56$ & $18.03$ & $0.16\pm0.17$\\ % $15.39\pm0.07$ & $15.23\pm0.16$ \\
J135224.06-252815.7 & star  & $18.13\pm0.02$ & $18.08\pm0.04$ &
$19.74\pm0.03$ & $18.49$ & $17.97$ & $0.22\pm0.23$\\ % $15.44\pm0.06$ & $15.22\pm0.22$ \\
J150144.61-115936.6 & star  & $18.12\pm0.03$ & $18.15\pm0.09$ &
$19.97\pm0.01$ & $18.49$ & $17.84$ & $0.14\pm0.19$\\ % $15.08\pm0.05$ & $14.94\pm0.18$ \\
J210326.83-261132.5 & star  & $18.20\pm0.03$ & $18.00\pm0.03$ &
$19.85\pm0.02$ & $18.58$ & $18.01$ & $0.48\pm0.28$\\ % $15.78\pm0.18$ & $15.30\pm0.20$ \\
\hline
J093429.75-052420.7 & star  & $18.18\pm0.02$ & $17.94\pm0.02$ &
$20.30\pm0.11$ & $18.41$ & $17.94$ & $0.26\pm0.23$\\ % $15.66\pm0.07$ & $15.40\pm0.22$ \\
J112908.13-300227.2 & star  & $17.98\pm0.04$ & $17.47\pm0.01$ &
$19.70\pm0.01$ & $18.27$ & $17.79$ & $-0.02\pm0.15$\\ % $14.95\pm0.04$ & $14.97\pm0.14$ \\
\hline
%J035555 & NA & $17.31\pm0.04$ & $16.56\pm0.03$ &
%$19.08\pm0.01$ & $17.36\pm0.01$ & $16.66\pm0.01$ & $13.47\pm0.02$ & $12.17\pm0.01$
\end{tabular}
\end{center}
\end{table*}

\section{DISCUSSION AND OUTLOOK}\label{Section_discussion}

This paper presents the first step in searching for high-redshift quasars with the SkyMapper Southern Survey. We first aimed for purity rather than completeness, and select candidate lists with relatively low contamination by cool stars. We thus employ the help of the deeper Pan-STARRS1 $g/r$-bands; however, the SkyMapper DR2 planned for August 2018 will provide deep photometry in all bands across the Southern declination range, although full coverage of the sky at full depth will only be achieved when the observations are completed in 2020.

We use a rather strict criterion of $g>21$, even though a few $z>4.5$ quasars are known to have $g<21$ in Pan-STARRS1 \citep[e.g.][]{Storrie01, Paris17}. All known $z>4.8$ quasars are fainter than $g=21.8$, and loosening the $g$-band limit rapidly increases the contamination by stars. Both of our newly discovered quasars are fainter than $g=21.5$. We also chose to use the WISE measurements of w1mag instead of w1mpro because contamination increases with w1mpro, and we selected the candidates to be isolated point sources.

Our two newly discovered $z\sim5$ quasars, SMSS J013539.27-212628.4 and SMSS J093032.58-221207.7, are bright, with $i\approx 18$ mag in the SkyMapper $i$-band; they are now two of the dozen brightest known $z>4.5$ quasars.

We note that one of our newly discovered quasars, SMSS J093032-221208, is outside of the selection box proposed by \cite{Wang16}, which misses a fraction of quasars that are blue in W1$-$W2 colour. Here, our MIR colour criteria that are sensitive to SED curvature are helpful as they include a wide redshift range, while the high-z selection is achieved by the optical broadband criteria. However, the star and quasar distributions are more blended than the previous criteria in \cite{McGreer13} and \cite{Wang16}. Thus, we adopted a strict selection box in the optical colours (Fig.~\ref{Figure_optical_criteria}) and a lenient one in the z-MIR colour (Fig.~\ref{Figure_MIR_criteria}). 

The current selection boxes are aimed at high purity and use very strict optical cuts. Loosening the optical criteria will increase star contamination faster than it increases completeness of the quasar population. This process will, however, become realistic now that proper motions for faint objects are available from Data Release 2 of Gaia \citep{Gaia16, GaiaDR2}. After finishing this work, Gaia DR2 became available and delivered proper motion data for virtualy all celestial objects to the depth of the candidates considered here. We found that all stars we identified had $>2\sigma$-significant proper motions, in excess of 1~milli-arcsec per year, while the true quasars at $z>4.5$ and $i<18.2$ have no significant proper motion. Gaia will not provide a route to a contamination-free selection of quasars, but will certainly reduce the contamination by a factor of $>10$.

%\begin{figure}
%\begin{center}
%\includegraphics[width=1.0\linewidth]{J035555.pdf}
%\caption{The spectrum of 035555-133033.}
%\label{Figure_unknown}
%\end{center}
%\end{figure}

\begin{acknowledgements}
This research was conducted by the Australian Research Council Centre of Excellence for All-sky Astrophysics (CAASTRO), through project number CE110001020.
ZL acknowledges financial support for a pre-PhD project at the ANU node of CAASTRO. ZL also thanks Yunyin Huang at Sun Yat-Sen University and Zhenzi He at Peking University for support on network access for TAP queries from China.
The national facility capability for SkyMapper has been funded through ARC LIEF grant LE130100104 from the Australian Research Council, awarded to the University of Sydney, the Australian National University, Swinburne University of Technology, the University of Queensland, the University of Western Australia, the University of Melbourne, Curtin University of Technology, Monash University and the Australian Astronomical Observatory. SkyMapper is owned and operated by The Australian National University's Research School of Astronomy and Astrophysics. The survey data were processed and provided by the SkyMapper Team at ANU. The SkyMapper node of the All-Sky Virtual Observatory (ASVO) is hosted at the National Computational Infrastructure (NCI). Development and support the SkyMapper node of the ASVO has been funded in part by Astronomy Australia Limited (AAL) and the Australian Government through the Commonwealth's Education Investment Fund (EIF) and National Collaborative Research Infrastructure Strategy (NCRIS), particularly the National eResearch Collaboration Tools and Resources (NeCTAR) and the Australian National Data Service Projects (ANDS). 
This work uses data products from the Wide-field Infrared Survey Explorer, which is a joint project of the University of California, Los Angeles, and the Jet Propulsion Laboratory/California Institute of Technology, funded by the National Aeronautics and Space Administration. AllWISE makes use of data from WISE, which is a joint project of the University of California, Los Angeles, and the Jet Propulsion Laboratory/California Institute of Technology, and NEOWISE, which is a project of the Jet Propulsion Laboratory/California Institute of Technology. WISE and NEOWISE are funded by the National Aeronautics and Space Administration.
This work has made use of data from the European Space Agency (ESA) mission {\it Gaia} (\url{https://www.cosmos.esa.int/gaia}), processed by the {\it Gaia} Data Processing and Analysis Consortium (DPAC, \url{https://www.cosmos.esa.int/web/gaia/dpac/consortium}). Funding for the DPAC has been provided by national institutions, in particular the institutions participating in the {\it Gaia} Multilateral Agreement.
\end{acknowledgements}

\begin{appendix}

\section{Queries for candidate lists}\label{appendix}

We selected candidates from the SkyMapper online database with the following ADQL Queries; the common section of both versions is:
\begin{verbatim}
SELECT m.*, w1mag, w1sigm, w2mag,
  w2sigm, cc_flags, var_flg, ngrp, 
  best_use_cntr, allwise.ph_qual, gmeanpsfmag, 
  gmeanpsfmagerr, rmeanpsfmag, rmeanpsfmagerr,
  imeanpsfmag, imeanpsfmagerr, zmeanpsfmag,
  zmeanpsfmagerr from dr1.master m
join ext.allwise on (allwise_cntr=cntr)
join ext.ps1_dr1 on (ps1_dr1_id=objid) 

WHERE u_psf is null and v_psf is null
and g_psf is null and r_psf is null
and i_psf between 16 and 18.2 
and z_psf-z_petro<0.35 
and abs(i_psf-imeanpsfmag)<0.4 
and abs(m.glat)>20 and m.prox>10 
and allwise_dist<2 and nb=1 and flags<4 
and nimaflags=0 and nch_max=1 
and ebmv_sfd<0.1 and ext_flg = 0 
and gmeanpsfmagerr>0 and rmeanpsfmagerr>0
and imeanpsfmagerr>0 and zmeanpsfmagerr>0
\end{verbatim}

Query addition for the SkyMapper version:
\begin{verbatim}
AND ((gmeanpsfmag<0 or gmeanpsfmag>21.0))
and (rmeanpsfmag-i_psf>0.50)
and i_psf-z_psf
  <0.40*(rmeanpsfmag-i_psf)-0.40
and (i_psf-z_psf<0.40)
and (z_psf-w1mag-w1mag+w2mag) 
  <-1.25*(i_psf-z_psf-z_psf+w1mag)-0.70
and (i_psf-z_psf-z_psf+w1mag)<-2.00
\end{verbatim}

Query addition for the Pan-STARRS1 version:
\begin{verbatim}
AND ((gmeanpsfmag<0 or gmeanpsfmag>21.0))
and (rmeanpsfmag-imeanpsfmag>0.50)
and (imeanpsfmag-zmeanpsfmag<
 0.45*(rmeanpsfmag-imeanpsfmag)-0.10)
and (imeanpsfmag-zmeanpsfmag<0.72)
[and iPSFMag-iKronMag<0.05]
\end{verbatim}

\end{appendix}

\bibliographystyle{pasa-mnras}

\end{document}